\documentstyle[prl,aps,twocolumn,epsfig]{revtex}

\begin{document}

\draft

\title{An NMR analog of the quantum disentanglement eraser}
\author{G. Teklemariam$^{\dagger}$, E. M. Fortunato$^{\ddagger}$, 
M. A. Pravia$^{\ddagger}$, T. F. Havel$^{\ddagger}$, D. G. Cory$^{\ddagger}$}

\address{$^{\dagger}$Department of Physics, MIT \\
$^{\ddagger}$Department of Nuclear Engineering, MIT}

\maketitle

\newcommand{\ket}[1]{$\vert${#1}$\rangle$}
\newcommand{\mket}[1]{\vert{#1}\rangle}
\newcommand{\mbra}[1]{\langle{#1}\vert}
\newcommand{\tfrac}[2]{{\textstyle\frac{#1}{#2}}}
\def\sig1{\sigma_1}
\def\sig2{\sigma_2}
\def\sig3{\sigma_3}
\def\isig1{\iota\sigma_1}
\def\isig2{\iota\sigma_2}
\def\isig3{\iota\sigma_3}
\def\up{|\uparrow\,\rangle}
\def\dn{|\downarrow\,\rangle}
\def\upd{\langle\, \uparrow |}
\def\dnd{\langle\, \downarrow |}
\def\beqn{\begin{equation}}
\def\eeqn{\end{equation}}
\def\beqnar{\begin{eqnarray}}
\def\eeqnar{\end{eqnarray}}
\oddsidemargin -0.2in
\evensidemargin -0.2in
\def\ba{\begin{array}}
\def\ea{\end{array}}

\begin{abstract}
We report the implementation of a three-spin
quantum disentanglement eraser on a
liquid-state NMR quantum information processor. 
A key feature of this experiment was its use of pulsed
magnetic field gradients to mimic projective measurements.
This ability is an important step towards the development
of an experimentally controllable system which can simulate
{\em any\/} quantum dynamics, both coherent and decoherent.
\end{abstract}


\def\beqn{\begin{equation}}
\def\eeqn{\end{equation}}
\def\beqnar{\begin{eqnarray}}
\def\eeqnar{\end{eqnarray}}
\oddsidemargin -0.2in
\evensidemargin -0.2in
\newcommand{\mb}[1]{\mbox{\boldmath{$#1$}}}
\def\ba{\begin{array}}
\def\ea{\end{array}}
\newcommand{\wdg}{\! \wedge \!}
\newcommand{\crs}{\! \times \!}
\newcommand{\scp}{\! \ast \!}
\newcommand{\dt}{\! \cdot \!}
\newcommand{\etal}{{\em et al. }}
\newcommand{\eqn}[1]{(\ref{#1})}


One of the most intriguing effects in quantum mechanics
is the ``quantum eraser'' \cite{Jaynes,Scul,Kwiat,Haroche}.
Given an ensemble of identically prepared quantum systems,
this effect is described by the {\em loss or gain} of
interference in a subensemble that is determined by
the outcome of the measurement of {\em one or the other}
of a pair of noncommuting binary observables, respectively.
Thus a quantum eraser demonstrates the principle of complementarity
without making use of the corresponding uncertainty relation.
Quantum erasers have previously been demonstrated by optical
\cite{KwiatEtAl} as well as atom \cite{DurrEtAl} interferometry.
In this Letter we use liquid-state NMR spectroscopy on pseudo-pure
states \cite{Price} to demonstrate a novel ``disentanglement'' eraser,
due to Garisto and Hardy \cite{Garisto}, in which not only interference,
but also entanglement, is lost or gained in a subensemble.
An analogous quantum erasure procedure operating on a pair
of Bell states has recently been used by Zeilinger's group
to prepare an entangled three-photon state \cite{Zeilinger}.

An important goal of our group is to design and build increasingly
more powerful experimentally controllable devices capable
of precisely simulating the {\em dynamics\/} of any quantum
system with an equal or smaller Hilbert space dimension.
Previously we have addressed the issues of coherent control
\cite{CoryEtAl} and pseudo-pure state preparation \cite{Havel1},
and we are now developing methods for non-unitary quantum operations.
The disentanglement eraser is of particular interest in this regard,
because it allows us to show that NMR on pseudo-pure states is
capable of reproducing even the decoherent dynamics associated with
{\em strong} (projective) measurements on the members of the ensemble,
which are needed to create or destroy entanglement in this eraser
(cf.~\cite{Havel1}).

It should be understood that the density matrices of the
highly mixed macro-states involved in liquid-state
NMR experiments can always be rationalized in terms
of ensembles of disentangled micro-states \cite{Caves}.
Consequentially, the ensemble-average observations on
pseudo-pure states reported here do not prove the existence
of the corresponding entangled micro-states in the sample.
Nevertheless, because pseudo-pure states provide an equivalent
representation of the underlying quantum {\em dynamics},
our experiments created exactly the same ensemble-average
coherences that would have been observed if the same operations
had been applied to the corresponding pure state ensemble,
and this is sufficient for our purposes.

In the disentanglement eraser two of the spins (qubits) in
a GHZ (Greenberger-Horne-Zeilinger) state \cite{GHSZ,Mermin}
are regarded as the components of a Bell state labeled by the
state of an additional ``ancilla'' spin \#1 (left-most), i.e.
\beqn
\mket{\psi_{GHZ}} = \tfrac1{\sqrt2}
(\mket{0}\mket{00} + \mket{1}\mket{11}) ~.
\eeqn
Assuming that the computational basis corresponds to
the eigenvectors of the $\sigma_z$ spin $1/2$ operator,
a projective measurement of the ancilla along $z$
yields a mixture of separable states $\mket{00}$
and $\mket{11}$ labeled by the ancilla spin.
This corresponds to the ensemble
\beqn
\rho_z = E^1_+\mket{00}\mbra{00}+E^1_-\mket{11}\mbra{11},
\eeqn
where $E_{\pm}^1 = \frac{1}{2}(I \pm \sigma_z^1)$ expresses the density
matrices $|0\rangle\langle0|\otimes \sigma_1 \otimes \sigma_1 = E_+^1$ and
$|1\rangle\langle1|\otimes \sigma_1 \otimes \sigma_1 = E_-^1$ in terms of the
Pauli matrix $\sigma_z^1 \equiv \sigma_z \otimes \sigma_1 \otimes \sigma_1$,
where $\sigma_1 = \mket0\mbra0+\mket1\mbra1$ and $I = \sigma_1 \otimes
\sigma_1 \otimes \sigma_1$ is the $8 \times 8$ identity matrix.

Alternatively, expressing the GHZ state in terms of the $x$ eigenstates
of the ancilla spin and the Bell states of the other two spins yields
\beqn
\mket{\psi_{GHZ}}
=\tfrac1{\sqrt2}(\mket{x_+}\mket{\phi_+}+\mket{x_-}\mket{\phi_-}),
\eeqn
where $\mket{x_\pm} = (\mket0 \pm \mket1) / \sqrt2$ and
$\mket{\phi_{\pm}}=(\mket{00}\pm \mket{11})/\sqrt{2}$.
Thus a projective measurement along the $x$-axis followed
by a rotation of the ancilla back to $z$ gives
\beqn
\rho_x = E^1_+\mket{\phi_+}\mbra{\phi_+} + E^1_-\mket{\phi_-}\mbra{\phi_-}.
\eeqn
This is a mixture of complementary Bell states
each labeled by the state of the ancilla.
Note that the partial trace over the ancilla
in $\mket{\psi_{GHZ}}\mbra{\psi_{GHZ}}$,
$\rho_z$ and $\rho_x$ are all equal to
$(\mket{00}\mbra{00} + \mket{11}\mbra{11})/2
= E_+^2 E_+^3 + E_-^2 E_-^3$, so that these
states can be distinguished only if the information
contained in the state of the ancilla is used.

These effects were demonstrated by liquid-state NMR
using as the qubits the three spin $1/2$ carbons
in a $^{13}$C-labeled sample of alanine
($C^1O_2^--C^2H(C^3H_3)-NH_3^+$) in deuterated water.
With decoupling of the protons \cite{Freeman},
this spin system exhibits a weakly coupled
spectrum corresponding to the Hamiltonian
\beqnar 
{\cal H}_{int}=\pi[\nu_1\sigma^1_z+\nu_2\sigma^2_z+\nu_3
\sigma^3_z \hspace{2.5cm} \nonumber \\
\hspace{1.5cm}+\tfrac{1}{2}(J_{12}\sigma^1_z\sigma^2_z
+J_{23}\sigma^2_z\sigma^3_z+J_{13}\sigma^1_z\sigma^3_z)],
\eeqnar
where the $\nu$'s are Larmour frequencies and
the $J$'s the spin-spin coupling constants in Hertz.
The experiments were carried out on a Bruker AVANCE-300
spectrometer in a field  of roughly 7.2 Tesla,
where the resonant frequency of the
second carbon is 75.4713562 MHz.
The frequency shifts of the other carbons with respect to the
second are 9456.5 Hz for the first one and -2594.3 Hz for the third,
while the coupling constants are $J_{12}=53.7$,
$J_{23}=34.6$ and $J_{13}=-1.4$ Hz.
The $T_1$ relaxation times for the three spins are $21$,
$2.5$ and $1.6$ s, while the $T_2$ times are $550$,
$420$ and $800$ ms, respectively.

The pseudo-pure ground state was prepared from the thermal
equilibrium state by the procedure summarized in Table 1,
which uses magnetic field gradients (denoted by $[\nabla]$)
to dephase off-diagonal elements of the density matrix
at strategic points along the way \cite{Price}.
Letting $\hat\rho_{eq} = \sigma_z^1 + \sigma_z^2 + \sigma_z^3$
be the traceless part of the equilibrium density
matrix (with all physical constants set to unity),
the first two transformations in the table yield the state
$({\scriptstyle \sqrt3/\sqrt{32}})\sigma_z^2 + (\sigma_z^1 + \sigma_z^3)
E_+^2\,$.
Spins $1$ and $3$ may then be transformed into the state
$({\scriptstyle \sqrt3/\sqrt{32}})(\sigma_z^1 + \sigma_z^3 +
\sigma_z^1\sigma_z^3)$
by the efficient two-spin pseudo-pure state preparation
procedure described in Ref.~\cite{Havel1} (Eq.~(47)),
yielding the three-spin pseudo-pure ground state
\beqn
\hat\rho_{ini} = \tfrac{\sqrt3}{\sqrt{32}} \left( E^1_+E^2_+E^3_+ -
\tfrac18 \right) \equiv \left( \mket{000}\mbra{000} - \tfrac18 \right) \,.
\eeqn

The logic network shown in Fig.~1 transforms
this state into the pseudo-pure GHZ state,
and then decohers the ancilla as indicated.
The GHZ state is obtained by rotating spin $2$
(since $J_{13} \ll J_{12\,}, J_{23}$)
to the $x$ axis in the rotating frame with a $\pi/2$
$y$-rotation $R^2_y(\pi/2) \equiv \exp(-i \sigma_y^2 \pi / 4)$,
and then using it as the control for a pair of c-NOT
(controlled-NOT \cite{Stean}) gates to the other two spins.
This pair of c-NOT's was implemented by the propagator
$N^{13|2} \equiv e^{i(\sigma^1_x-\sigma^3_x)E^2_-\pi/2}
= e^{i\sigma^1_xE^2_-\pi/2} e^{-i\sigma^3_xE^2_-\pi/2}$
(ensuring cancellation of the phases
$\pm i$ between the two factors).
The overall sequence of transformations
on the corresponding state vector is thus:
\begin{eqnarray}
\mket{000} &\stackrel{R^2_y(\pi/2)}{\begin{picture}(30,4)%
\put(0,2.5){\line(1,0){20}}\put(19,0){$\rightarrow$}\end{picture}}&
\mket0 (\mket0 + \mket1) \mket0 / \sqrt2 \nonumber \\
&\stackrel{N^{13|2}}{\begin{picture}(30,5)\put(0,2.5){%
\line(1,0){20}}\put(19,0){$\rightarrow$}\end{picture}}&
(\mket{000} + \mket{111}) / \sqrt2 \equiv \mket{\psi_{GHZ}}
\end{eqnarray}
Implementations of these operations in NMR
by RF (radio-frequency) pulse sequences may
be found in Refs.~\cite{Price,Havel1,Shy}.
The resulting pseudo-pure GHZ state is written
in product operator notation as \cite{Shy}
\begin{eqnarray}
\hat{\rho}_{GHZ}=\tfrac{\sqrt{3}}{4\sqrt{2}}(\sigma^1_z\sigma^2_z
+\sigma^2_z\sigma^3_z+\sigma^1_z\sigma^3_z \hspace{2.5cm}
\nonumber \\
 +\sigma^1_x\sigma^2_x\sigma^3_x
-\sigma^1_y\sigma^2_y\sigma^3_x-\sigma^1_x\sigma^2_y\sigma^3_y
-\sigma^1_y\sigma^2_x\sigma^3_y) \,,
\end{eqnarray}
and has previously been studied by NMR in Refs.~\cite{Rick,Laf}.

The coherences of $\rho_{GHZ}$ can be dephased,
exactly as they would be by strong measurements of
$\sigma_z$ on all the individual systems in the ensemble,
by means of magnetic field gradients similar to those used
to prepare the initial pseudo-pure state (cf.~\cite{Havel1}).
Specifically, a constant gradient $\nabla = \partial B_z / \partial z$
applied for a period $t$ along the static field
axis $z$ causes spin evolution under the Hamiltonian
$z \gamma \nabla \tfrac{1}{2} \sum^{3}_{j=1}  \sigma^j_z$,
where $\gamma$ is the gyromagnetic ratio of all the spins.
This multiplies each coherence $\rho_{k\ell}$ ($k\ne\ell$) with a
spatially dependent phase $\exp(-i \gamma m_{k\ell} \nabla z t / 2)$,
where $m_{k\ell}$ is the {\em coherence order\/} \cite{Sodickson}
(i.e.~the difference in the $z$-component of the angular momentum in units
of $\hbar$ between the $\mket{k}$ and $\mket\ell$ states \cite{Freeman}).
Thus after such a gradient pulse the density matrix averaged
over the sample volume satisfies $\rho_{k\ell} = 0$ for all
$k \ne \ell$ save for the zero quantum coherences ($m_{k\ell} = 0$).
Because only one spin is dephased in the eraser experiments,
only single quantum coherences are of consequence.

This dephasing operation was made specific to
those coherences involving transitions of the ancilla
spin $1$ by applying a $\pi$ pulse to the other two spins,
after which a second gradient pulse of the same amplitude
and duration ``refocuses'' all the other coherences.
At the same time it is necessary to also refocus
the evolution under the internal Hamiltonian
using $\pi$ pulses selective for single spins.
A sequence of RF and gradient pulses which
accomplishes this is (in temporal order):
\begin{eqnarray}
   P^1_z &=&
   \left[ \nabla \right]_z -
   \left[ \pi \right]^2_{x} -
   \left[ \nabla \right]_z -
   \left[ \pi \right]^{2,3}_{x} -
\nonumber \\ &&
   \left[ \nabla \right]_z -
   \left[ \pi \right]^2_{-x} -
   \left[ \nabla \right]_z -
  \left[ \pi \right]^{2,3}_{-x}
\end{eqnarray}
The corresponding effective (average) propagator
is simply $e^{-i\nabla z\sigma^1_z/2}$.
This dephases the ancilla spin in the same way as would a strong
measurement of $\sigma_z^1$ on every member of the ensemble.
To dephase the ancilla in the same way as would a strong measurement of
$\sigma_x^1$, one need only rotate the ancilla to the $z$-axis with a
$\pi/2$ $y$-rotation $R^1_{-y}(\pi/2)$, as follows:
\beqn
P^1_x = \left[ \pi/2 \right]_{-y}^1 - P^1_z
\eeqn
The ancilla is left along $z$ for subsequent tomography.

The results of $P^1_z$ and $P^1_x$ applied to $\hat\rho_{GHZ}$ are
\begin{eqnarray}
\hat{\rho}_{GHZ}\stackrel{P^1_z}{\longrightarrow}
\tfrac{\sqrt{3}}{4\sqrt{2}}(\sigma^1_z\sigma^2_z
+\sigma^2_z\sigma^3_z+\sigma^1_z\sigma^3_z),
\hspace{.5cm}\\
\hat{\rho}_{GHZ}\stackrel{P^1_x}{\longrightarrow}
\tfrac{\sqrt{3}}{4\sqrt{2}}(\sigma^2_z\sigma^3_z
+\sigma^1_z\sigma^2_x\sigma^3_x-\sigma^1_z\sigma^2_y\sigma^3_y).
\end{eqnarray}
These states were confirmed by full tomography \cite{Ike}.
Because only the single quantum ($m_{k\ell} = 1$)
coherences give rise to observable (dipolar) magnetization,
it is necessary to collect spectra not only following
the dephasing operation, but also following additional
$\pi/2$ pulses selective for single spins, to rotate
the $m_{k\ell}=0$ and $m_{k\ell}>1$ coherences,
as well as the populations (diagonal elements),
into observable single quantum coherences.
Tomography was performed at the points of the
procedure indicated in Fig.~1; the real parts
of these four density matrices are shown in Fig.~2
(the imaginary parts were essentially zero).

The overall precision of quantum information transmission was
quantified by an extension of Schumacher's fidelity \cite{Schu},
which takes into account not only systematic errors,
but also the net loss of magnetization due to random errors.
This measure, called the {\em attenuated correlation}, is given by
\begin{eqnarray}
c(\hat{\rho}^{exp}) = \frac{Tr(\hat{\rho}^{the} \hat{\rho}^{exp})}
{Tr(\hat{\rho}^{the} \hat{\rho}^{the})}.
\end{eqnarray}
Here, $\hat{\rho}^{the}$ is the measured
pseudo-pure ground state $\hat\rho_{ini}^{exp}\,$,
transformed on a computer by the same sequence
of unitary and non-unitary (measurement)
operations to which it was subjected on the
spectrometer to get $\hat{\rho}^{exp}\,$.
Note that, since $Tr(\hat\rho^{exp}\hat\rho^{exp}) \le
Tr(\hat\rho^{the}\hat\rho^{the})$, the Cauchy-Schwarz
inequality implies that $-1 \le c(\hat\rho^{exp}) \le 1$.

The values of the correlation for each
of the four tomographic readouts were
$c(\hat\rho_{ini}^{exp}) = 1$ (by definition),
$c(\hat\rho_{GHZ}^{exp}) = 0.88$, $c(\hat\rho_z^{exp})
= 0.92$ and $c(\hat\rho_x^{exp}) = 0.93$. Although not 
included here for brevity, tomography on the state
$\mket{0}(\mket{00}+\mket{11})/\sqrt2$ yields
an attenuated correlation of $0.93$, showing that spins
2 and 3 were entangled before the GHZ state was created.
The increases in $c$ are not unexpected, since the additional
$\pi$ and gradient pulses needed to mimic measurements on
$\rho_{GHZ}$ are easily implemented with high precision,
and the tomographic errors are estimated at $\pm5$\%.
The leading candidates for the loss of correlation are
pulse imperfections arising from RF field inhomogeneity,
less than perfect RF pulse calibrations, and relaxation.
The total time before data collection in
the complete experiments was ca.~$65$ ms;
the time required to prepare the GHZ state
from the initial state was $21$ ms.
Since the $T_2$ relaxation times of the spins varied from $400-800$ ms,
the net loss of magnetization due to relaxation in going
from $\hat\rho_{ini}$ to $\hat\rho_{GHZ}$ was $3-5$\%.
Thus, the additional loss due to pulse imperfections etc.~was
about another 5\% or so, confirming the high precision of the
strongly modulated pulse sequences used in these experiments.

In conclusion, we have used a three-spin liquid-state NMR
quantum information processor to obtain a high-precision
implementation of the dynamics, both coherent and decoherent,
underlying Garisto and Hardy's ``disentanglement eraser'',
and have found that the experimental results confirm the
theoretically predicted conditional expectation values.
This shows that we can judiciously and selectively render phase
information macroscopically inaccessible in a way that precisely
mimics the decoherence attendant on strong measurements.
It should be noted that during this dephasing operation
all interactions among the spins were refocused,
and that only the {\em macroscopically accessible\/}
information contained in the ancilla spin due to its
earlier interactions with the other two was changed.
This was nevertheless sufficient to convert the net
triple-quantum coherence in $\rho_{GHZ}$ into a pair
of double-quantum coherences, conditional on the
state of the ancilla representing this information.
Unlike previous eraser implementations, it was not necessary
to explicitly read out this information in each member
of the ensemble in order to see the conditional coherence,
because this was done for us by the coupling of the ancilla
to the other two spins while the spectra were being measured.
This ability to convert macroscopic correlations into
one another via well-defined microscopic (molecular)
interactions is the essence of ensemble quantum computing.

We thank L. Viola and S. Somaroo for helpful discussions.
This work was supported by the U.S. Army Research Office under grant
number DAAG 55-97-1-0342 from the Defense Advanced Research
Projects Agency.

Correspondence should be addressed to DGC (email:{\it dcory@mit.edu}).

\begin{figure*}
\setlength{\unitlength}{.35cm}
\begin{picture}(10,7.5)
\put(-.25,5.85){$E^3_+$}
\put(-.25,3.35){$E^2_+$}
\put(-.25,.85){$E^1_+$}
\put(4.,6.75){\vector(0,-1){.75}}
\put(1.5,7){Tomography}
\put(-1.,-.75){$\rho_{ini}$}
\put(2.,6){\line(2,0){14}}
\put(2.,3.5){\line(2,0){4.25}}
\put(2.,1){\line(2,0){4}}
\put(6.25,2.75){\framebox(1.5,1.5){$\frac{\pi}{2}]_y$}}
\put(7.75,3.5){\line(2,0){10}}
\put(8.5,6){\line(2,0){1.5}}
\put(13.,-.75){$\rho_{GHZ}$}
\put(10.,3.5){\line(0,2){2.5}}
\put(10.,3.5){\circle*{.45}}
\put(9.75,5.775){\sf x}
\put(12.5,1){\line(0,2){2.5}}
\put(12.5,3.5){\circle*{.45}}
\put(12.25,.75){\sf x}
\put(5.5,1){\line(2,0){7}}
\put(13.,3.5){\line(2,0){1}}
\put(13.,6){\line(2,0){1}}
\put(12.5,1){\line(2,0){1}}
\put(14.5,7.75){\vector(0,-1){1.75}}
\put(11.5,8.25){Tomography}
\put(13.5,1){\line(2,0){2.25}}
\put(15.75,.15){\framebox(4.2,2.15)[t]{Decoher}}
\put(15.75,.1){\framebox(4.2,2)[lb]{along x/z}}
\put(13.5,3.5){\line(2,0){5.5}}
\put(13.5,6){\line(2,0){5.5}}
\put(19.,3.5){\line(2,0){1}}
\put(19.,6){\line(2,0){1}}
\put(19.5,6.75){\vector(0,-1){.75}}
\put(16.5,7){Tomography}
\put(20.5,-.75){$\rho_{x/z}$}
\put(20.,3.5){\line(2,0){1}}
\put(20.,6){\line(2,0){1}}
\end{picture}
\vspace{.3 in}
\caption{\small{Logic network for the disentanglement eraser.  
Initially, a pseudo-pure state on spins 1, 2 and 3 is created, 
$\rho_{ini}=\mket{000}\mbra{000}\equiv E^1_+E^2_+E^3_+$. 
A $\frac{\pi}{2}$ $y$-pulse is then applied to spin 2, followed by
two controlled-not (c-NOT) gates to create the GHZ state (see text).
Conditionality on the second spin being in the $\mket1$ state is
represented in the network by a filled circle on its time line.
Finally, the two complementary measurements, 
$\sigma^1_z$ and $\sigma^1_x$, are applied to spin 1. 
State tomography was performed to fully reconstruct
the density matrices at the positions indicated.}}
\end{figure*}
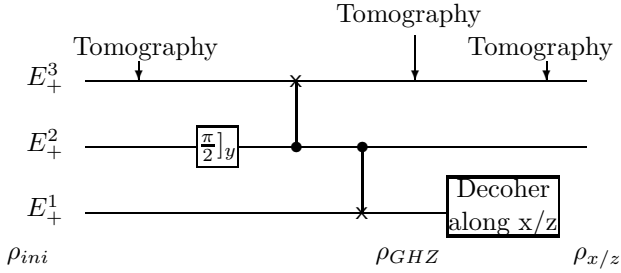

\begin{table}
\setlength{\unitlength}{.35cm}
\begin{picture}(10,10.5)
\put(4,10){Transformations}
\put(0,9.5){\line(2,0){15}}

\put(0.,7.5){1)}
\put(1.5,7.5){$[\nabla]e^{-\frac{i}{2}\cos^{-1}(\frac{\sqrt{3}}{4\sqrt{2}})
\sigma^2_x}$}

\put(0.,5.5){2)}
\put(1.5,5.5){$[\nabla]e^{-i\frac{\pi}{4}(\sigma^1_y+\sigma^3_y)E^2_-}$}

\put(0.,3.5){3)}
\put(1.5,3.5){$e^{i\frac{\pi}{4}\sigma^1_x}e^{-i\frac{\pi}{4}\sigma^1_z\sigma^2
_z}
e^{-i\frac{\pi}{4}(\sigma^1_y+\sigma^2_y)}e^{-i\frac{\pi}{4}\sigma^1_z\sigma^2_
z}
e^{-i\frac{\pi}{4}\sigma^2_x}$}

\put(0.,1.5){4)}
\put(1.5,1.5){$[\nabla]e^{i\frac{\pi}{12}(\sigma^2_y+\sigma^3_y)}
e^{-i\frac{\pi}{4}
\sigma^1_z\sigma^2_z}e^{-i\frac{\pi}{8}(\sigma^2_x+\sigma^3_x)}$}






\end{picture}
\caption{\small{This table shows the transformations used
to obtain the initial state from the thermal state, $\rho_{eq}$. 
}}
\end{table}

\begin{figure*}
  {\epsfxsize=3.in\epsfysize=7.1in\centerline{\epsffile{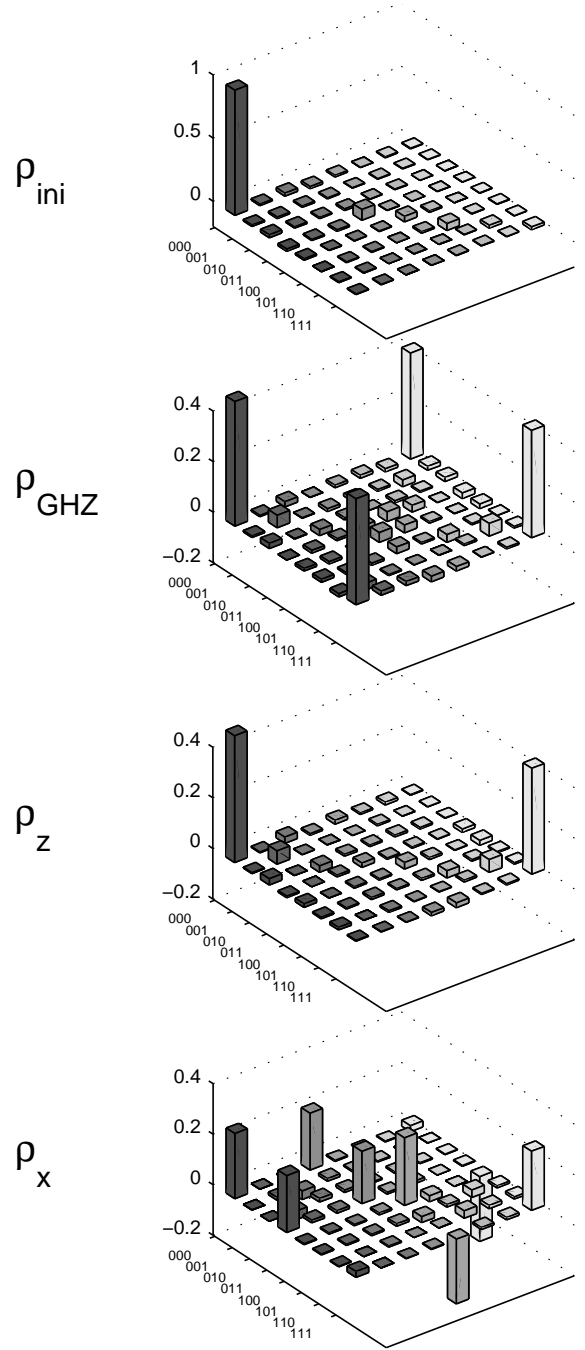}}}
\vspace{.5cm}
\caption{\small{Experimental density matrices reconstructed by tomography
(in normalized units). The rows are enumerated in the standard computational
basis, where for example 000 represents the state label $\mket{000}$.
Although not shown, the columns are similarly labeled with the leftmost end
representing $\mket{000}$ and the rightmost end representing $\mket{111}$.
$\rho_{ini}$ is the three-spin pseudo-pure ground state, and $\rho_{GHZ}$
is the pseudo-pure GHZ state. The last two plots are $\rho_{z\,}$,
which is $\rho_{GHZ}$ after decohering spin 1 about the $z$-axis,
and $\rho_{x\,}$, which is after decohering it about the $x$-axis.
 (Note:
$\rho_{GHZ}$, $\rho_{z}$ and $\rho_{x}$ have
been magnified by a factor of two for clarity). 
An amount of identity, chosen to optimize the input projection,
was added to all experimentally measured density matrices.
} }
\end{figure*}

\end{document}